\documentclass[10pt, conference]{IEEEtran}
\ifCLASSINFOpdf
   \usepackage[pdftex]{graphicx}
\else
\fi
\usepackage{url}

\usepackage{enumitem} 
\usepackage{multirow}
\usepackage{xcolor}
\usepackage{colortbl}
\usepackage{fancyvrb}
\usepackage{xspace}
\usepackage{flushend}
\usepackage{listings}
\usepackage{etoolbox}

\makeatletter
\preto\lstlisting{\def\@captype{table}}
\makeatother

\usepackage[colorlinks   = true,
citecolor    = blue,
linkcolor    = black,
urlcolor	 = black]{hyperref}

\usepackage{lipsum}

\newcommand{\secref}[1]{Section~\ref{#1}\xspace}
\newcommand{\figref}[1]{Figure~\ref{#1}\xspace}
\newcommand{\tabref}[1]{Table~\ref{#1}\xspace}
\newcommand{\lstref}[1]{Listing~\ref{#1}\xspace}
\newcommand{\ie}{\emph{i.e.},\xspace}
\newcommand{\eg}{\emph{e.g.},\xspace}
\newcommand{\etal}{\emph{et al.}\xspace}

 \usepackage{amssymb}                    
 \usepackage{ifthen}
 \newboolean{showcomments}
 \setboolean{showcomments}{true}  
 \ifthenelse{\boolean{showcomments}}
 {\newcommand{\nnote}[2]{
 		\fbox{\bfseries\sffamily\scriptsize#1}
 		{\sf\small$\blacktriangleright$\textit{#2}$\blacktriangleleft$}
 	}
 }
 {\newcommand{\nnote}[2]{}
 	
 }

\newcommand{\maven}{\textsc{Maven}}
\newcommand{\differname}{\textsc{BuildDiff}}

\newcommand\reppackagemarker[1]{\textcolor{blue}{}} 

\begin{document}
%
\title{Extracting Build Changes with \differname{}}



\author{\IEEEauthorblockN{Christian Macho}
	\IEEEauthorblockA{Software Engineering Research Group\\
		University of Klagenfurt\\
		Klagenfurt, Austria\\
		Email: christian.macho@aau.at}
	\and
	\IEEEauthorblockN{Shane McIntosh}
	\IEEEauthorblockA{Dept. of Electrical and Computer Eng. \\
		McGill University\\
		Montr\'{e}al, Canada\\
		Email: shane.mcintosh@mcgill.ca}
	\and
	\IEEEauthorblockN{Martin Pinzger}
	\IEEEauthorblockA{Software Engineering Research Group\\
		University of Klagenfurt\\
		Klagenfurt, Austria\\
		Email: martin.pinzger@aau.at}
}


%


\maketitle
\begin{abstract}
Build systems are an essential part of modern software engineering projects. As software projects change continuously, it is crucial to understand how the build system changes because neglecting its maintenance can lead to expensive build breakage. Recent studies have investigated the (co-)evolution of build configurations 
and reasons for build breakage, 
but they did this only on a coarse grained level. 

In this paper, we present \differname{}, an approach to extract detailed build changes from \maven{} build files and classify them into 95 change types. 
%
In a manual evaluation of 400 build changing commits, we show that \differname{} can extract and classify build changes with an average precision and recall of 0.96 and 0.98, respectively. We then present two studies using the build changes extracted from 30 open source Java projects to study the frequency and time of build changes. 
The results show that the top 10 most frequent change types account for 73\% of the build changes. Among them, changes to version numbers and changes to dependencies of the projects occur most frequently. Furthermore, our results show that build changes occur frequently around releases. 

With these results, we provide the basis for further research, such as for analyzing the (co-)evolution of build files with other artifacts or improving effort estimation approaches. Furthermore, our detailed change information enables improvements of refactoring approaches for build configurations and improvements of models to identify error-prone build files.
\end{abstract}

\begin{IEEEkeywords}
Maintenance, Build Systems, Software Quality
\end{IEEEkeywords}

%
\IEEEpeerreviewmaketitle

%
%
%
%


\newcommand{\RQOne}{Which build change types occur the most frequently?}
\newcommand{\RQTwo}{When are build changes recorded?}

\newcommand{\numberofprojects}{30}

\section{Introduction}
Large software projects use build tools, such as \textsc{Maven}, \textsc{Gradle}, or \textsc{Ant}, to automate the assembling and testing process of their software products. The configuration of such build systems can often be complex \cite{McIntosh2012Evolution}, which also complicates their maintenance. Seo \etal \cite{seo2014programmers} showed that up to 37\% of builds at Google fail, stating neglected build maintenance as the most frequent cause. The development team is then blocked and obliged to fix the build first. Kerzazi \etal \cite{kerzazi2014automated} found a similar ratio of up to 18\% of build breakage and estimated the total costs for the breakages in their study to be more than 336 man-hours. 

As a software system evolves, changes are applied to the source code. Developer teams also need to maintain the build and hence, subsequent changes need to be applied to the build configuration. Adams \etal \cite{Adams2008Evolution} and McIntosh \etal~\cite{McIntosh2012Evolution} found evidence of a co-evolutionary relationship between source and build code. Hence, omitting changes to the build configuration that are needed to remain synchronized with the source code, can lead to build breakage. To that end, it is important to know when build changes should be applied. McIntosh \etal \cite{McIntosh2014Mining} and Macho \etal \cite{Macho2016Predicting} studied this problem and found that their models can predict whether a source code change should have an accompanying build change. However, both studies lack detailed information about the type of build change that is needed.

In this study, we investigate changes to the build configuration in detail. We are interested in which types of changes are typically made to the build configuration and when they are performed. Prior studies \cite{McIntosh2014Mining} consider the build configuration to be changed if the build file changes but do not investigate the detailed type of the change. We claim that a more detailed view on build configuration changes can improve studies of the build system and its configuration. Thus, we introduce \textit{\differname}, an approach to extract build changes from \maven{} build files. Our approach is inspired by ChangeDistiller \cite{Fluri2007ChangeDistilling,Gall2009ChangeAnalysis}, which extracts source code changes from Java source files. To the best of our knowledge, we are the first to investigate build changes in this granularity. We also propose a taxonomy of build changes consisting of 95 build change types and 5 categories that our approach can extract. We evaluate \differname{} in a manual investigation of 400 randomly selected build changing commits and find that it yields an average precision and recall of 0.96  and 0.98, respectively.

Armed with an approach to extract build changes from \maven{} build files, we extract build changes from \numberofprojects{} open source Java projects from different vendors, of different sizes, and with different purposes. We study the extracted data in two ways. First, we study the frequency of build changes. We explore which change types are the most frequent ones and which change types are rarely applied. We analyze the frequencies also in terms of change categories. Second, we study the time at which build changes have been recorded and investigate whether build changes are equally distributed over a project or if there are periods in the projects where they occur more frequently. 
With the two studies, we address the following two research questions:
\begin{itemize}[leftmargin=1.12cm] 
	\item[\textbf{(RQ1)}] \textbf{\RQOne} \\ 
	 The most frequent build change type is PARENT\_VERSION\_UPDATE followed by PROJECT\_\-VERSION\_\-UPDATE, and DEPENDENCY\_\-INSERT. The most frequent build change category is \texttt{General Changes}, directly followed by \texttt{Dependency Changes}, and \texttt{Build Changes}. Changes to version properties and to dependencies can be found frequently among the top 10 most frequent change types. The top 10 change types account for 73\% of all of the changes.
	\item[\textbf{(RQ2)}] \textbf{\RQTwo} \\ Build changes are not equally distributed over the projects' timeline. There are particular phases that contain significantly more build changes than others. We observe that especially around releases, the frequency of build changes is high.
	
\end{itemize}

This work makes the following contributions: 
(1) an approach and a corresponding prototype implementation to extract fine-grained build changes from \maven{} build files, (2) a dataset containing historical build changes of \numberofprojects{} open source Java projects, (3) an evaluation of the performance of this tool, (4) two empirical studies of the frequency and time of build changes, and (5) a replication package that contains supplementary material.\footnote{\label{fn:repPackage}\url{https://figshare.com/articles/BuildDiff_Supplementary_Material/4786084}}



The remainder of the paper is organized as follows: \secref{sec:rw} situates the paper with respect to the related work. \secref{sec:buildDiff} presents our \differname{} approach. \secref{sec:dataPreparation} describes the data that we used for this study and evaluates the performance of \differname{}, and discusses its strengths and weaknesses. \secref{sec:changeFrequency} presents the first study on the frequency of build changes and \secref{sec:whenAndWhy} shows the second study on when build changes occur. \secref{sec:discussion} discusses the implications of our results and threats to validity. \secref{sec:conclusions} concludes the paper.

\section{Related Work}
\label{sec:rw}
\textbf{Build Maintenance.}
Related work on build maintenance includes the co-evolution of build systems with other artifacts of the development process. For instance, Adams \etal introduced MAKAO \cite{Adams2007DesignRecovery}, a framework for re(verse)-engineering build systems, and studied the co-evolution of the Linux build system \cite{Adams2008Evolution} using MAKAO. They found that the build system itself evolves and its complexity grows over time. Furthermore, they identified maintenance as the main factor for evolution. 
McIntosh \etal investigated the evolution of the ANT build system \cite{McIntosh2010Evolution} from a static and a dynamic perspective. They defined a metric for measuring the complexity of build systems and found that the complexity of ANT build files  evolves over time, too. In follow-up work, McIntosh \etal investigated Java build systems and their co-evolution with production and test code \cite{McIntosh2012Evolution}. The results of a large-scale study showed a relationship between build technology and maintenance effort~\cite{McIntosh2014LargeScale}. In addition to these studies, Hardt and Munson developed Formiga, a tool to refactor ANT build scripts~\cite{Hardt2015EmpiricalFormiga}~\cite{Hardt2013FormigaAntMaintenance}. 

Concerning the co-evolution of build configurations with other software artifacts, existing studies investigated models to predict build co-changes based on various metrics. For instance, McIntosh \etal \cite{McIntosh2014Mining} used code change characteristics to predict build co-changes within a software project. Xia \etal \cite{Xia2015CrossProject} extended this study by building a model for predicting build co-changes across software projects. Macho \etal~\cite{Macho2016Predicting} showed that they can improve both studies by using fine-grained source code changes. Furthermore, Xia \etal \cite{Xia2014BuildSystem} investigated missing dependencies in build files using link prediction. They showed that their algorithm outperforms state-of-the-art link prediction algorithms for this problem.

\textbf{Change Extraction.} 
Many previous studies used changes that were extracted from different versions of source files to investigate various aspects of the evolution of software systems. Miller \etal \cite{miller1985file} and Myers \etal \cite{myers1986ano} performed their studies on the level of text simply by counting the number of added or deleted lines of text. One advantage of these approaches is that they do not need a parser or a grammar to output the differences between the different versions of source files. However, one important shortcoming is that these approaches have difficulty mapping the changed lines of text to actual changes in source files, such as the change of the return type of a method or the addition of an else branch. 

Modern approaches overcome this issue by performing the differencing on the level of Abstract Syntax Trees (ASTs). For instance, Hashimoto and Mori \cite{hashimoto2008diff} developed Diff/TS, which is working on the raw AST created from parsing two versions of a source file. The most prominent approach in this area is ChangeDistiller of Fluri \etal \cite{Fluri2007ChangeDistilling}. Their approach extracts differences from two consecutive versions of a Java file and maps the differences to 48 change types \cite{Fluri2006Classifying}. Falleri \etal \cite{Falleri2014GumTree} improved the differencing algorithm by applying a combined method for matching equal subtrees and showed that their approach outperforms ChangeDistiller. 
Concerning build changes, D\'esarmeaux \etal \cite{desarmeaux2016mavenLifecyclePhases} mapped line-level changes to \maven~lifecycle phases and investigated the maintenance effort of each phase. They found out that the compile phase accounts for most of the maintenance.

In summary, we find that several approaches exist to study build maintenance, build systems, and their configuration. However, these studies are primarily based on coarse-grained metrics. Previous work showed that for programming languages, such as Java, the usage of a finer granularity of changes can help to improve prediction models \cite{Giger2011ComparingBugPrediction,Giger2012CodeChangePrediction,Giger:2012:MBP:2372251.2372285,Romano2011Predicting} or support the understanding of (co-)evolution \cite{Fluri2009coevolution}. To that extent, it is important to also investigate build changes on a fine-grained level. To the best of our knowledge, we are the first to present an approach to extract detailed changes from \maven{} build files.

\section{Extracting Build Changes with \differname{}}
\label{sec:buildDiff}
In this section, we describe our approach to extract build changes from build files. Currently, we focus on the extraction of \maven{} build files. \maven{} build files are named \texttt{pom.xml} following the naming convention of \maven{}. First, we define a taxonomy of build changes  and provide our rationale for the defined changes. Second, we describe \differname{}, our approach to extract build changes of two consecutive \maven{} build file revisions. 

\subsection{Taxonomy}
\label{ss:taxonomy}
\newcommand{\numberofchangetypes}{95}
\maven{} build files are specified using a special type of XML. Hence, we can easily read, parse, and transform their content into a tree that corresponds to the \maven{} schema\footnote{\url{http://maven.apache.org/xsd/maven-4.0.0.xsd}} that defines the various XML elements and attributes used for configuring a \maven{} build.
Having the content of a \maven{} build file represented as a tree, we then can use tree differencing algorithms, such as ChangeDistiller \cite{Fluri2007ChangeDistilling} or GumTree \cite{Falleri2014GumTree}, to extract differences between two build files. We use the modified version of the GumTree implementation of Dotzler \etal \cite{Dotzler2016moveoptimizeddiff} to extract edit operations that transform one tree into the other. In the remainder of the paper, we refer to this implementation as GumTree. We describe the extraction procedure in more detail in \secref{ss:approach}. 

Similar to ChangeDistiller, we defined the change types of our taxonomy based on the edit operations extracted by the tree differencing algorithm whereas the structure of the tree and its different elements correspond to the \maven{} schema. For defining the taxonomy, we started with the top level elements of the \maven{} schema and moved down the schema until we reached the bottom-most child elements. For each element (\ie XML tag), we defined change types for inserting (*\_INSERT), deleting (*\_DELETE), and updating (*\_UPDATE) that element. For some particular tags, such as \texttt{artifactId} and \texttt{groupId}, we only created the *\_UPDATE change type because they are mandatory for the definition of particular tags, such as \texttt{dependency}, and we assume that they are inserted and deleted with their parent tag. This is further described in detail in \secref{ss:approach}. The resulting taxonomy currently comprises \numberofchangetypes{} that we validated with two expert developers using \maven{}.

We also grouped the change types into categories. We retrieved the categories and the respective change type assignments by performing card sorting \cite{nielsen1995card}. First, we gave the list of change types to the two developers who validated the changes types separately and asked them to group the change types. Second, we asked the developers to assign names to the created groups. In a third step, we asked both developers to discuss their categories and assignments to arrive at a common categorization. If the developers assigned a change to different categories they discussed with one another to arrive at a consensus. 
\begin{table}[!t]
	\renewcommand{\arraystretch}{1.0}
	\caption{Excerpt of our Taxonomy of Build Changes}
	\label{tab:changesAndCategories}
	\centering
	\resizebox{\columnwidth}{!}{%
		\begin{tabular}{|l|l|l|}
			\hline
			Category & Change Types (Excerpt) \\\hline
			\multirow{3}{*}{Dependency Changes} & DEPENDENCY\_INSERT \\
			& DEPENDENCY\_VERSION\_UPDATE \\
			& MANAGED\_DEPENDENCY\_DELETE \\\hline
			
			\multirow{3}{*}{Build Changes}& PLUGIN\_INSERT     \\
			& PLUGIN\_CONFIGURATION\_UPDATE \\
			& TEST\_RESOURCE\_DELETE \\\hline
			
			\multirow{2}{*}{Team Changes}& DEVELOPER\_INSERT     \\
			& CONTRIBUTOR\_DELETE \\\hline
			
			\multirow{3}{*}{Repository Changes} & PLUGIN\_REPOSITORY\_INSERT    \\
			& DIST\_SNAPSHOT\_REPOSITORY\_UPDATE\\
			& REPOSITORY\_DELETE \\\hline
			
			\multirow{3}{*}{General Changes}& MODULE\_INSERT     \\
			& PARENT\_VERSION\_UPDATE \\
			& GENERAL\_PROPERTY\_DELETE \\\hline
		\end{tabular}%
	}
\end{table}

The two developers developed the following 5 categories: (1) \texttt{Dependency Changes} contain all changes that are related to dependencies of the project, (2) \texttt{Build Changes} cover the changes that directly affect or modify the build process, (3) \texttt{Team Changes} comprise all modifications to the list of team members, (4) \texttt{Repository Changes} hold changes that are performed to the distribution and repository locations, and (5) \texttt{General Changes} contain changes that are made to the general items of a \maven{} project.
\tabref{tab:changesAndCategories} shows an excerpt of the taxonomy with examples of change types for each category. The full taxonomy can be found online in the supplementary material.\textsuperscript{\ref{fn:repPackage}}

In the following, we provide two examples of frequently occurring build changes. The first example depicted in \lstref{lst:depVersionUpdate1} (old version of the build file) and \lstref{lst:depVersionUpdate2} (new version of the build file) shows a change of the version of a dependency to the \texttt{spring-core} library from \texttt{4.2.5.RELEASE} to \texttt{4.2.6.RELEASE}. This change is extracted and classified by \differname{} as DEPENDENCY\_VERSION\_UPDATE.

\vspace{0.5cm}
\begin{lstlisting}[frame=single,caption={Dependency Version Update - Old Version} ,label=lst:depVersionUpdate1]
<dependency>
 <groupId>org.springframework</groupId>
 <artifactId>spring-core</artifactId>
 <version>(*@\textcolor{red}{4.2.5.RELEASE}@*)</version>
</dependency>
\end{lstlisting}

\begin{lstlisting}[frame=single,caption=Dependency Version Update - New Version ,label=lst:depVersionUpdate2]
<dependency>
 <groupId>org.springframework</groupId>
 <artifactId>spring-core</artifactId>
 <version>(*@\textcolor{red}{4.2.6.RELEASE}@*)</version>
</dependency>
\end{lstlisting}
\vspace{0.5cm}
The second example depicted in \lstref{lst:plugInsert1} (old version of the build file) and \lstref{lst:plugInsert2} (new version of the build file) shows the insertion of the \texttt{maven-jar-plugin} plugin. This change is extracted and classified by \differname{} as PLUGIN\_INSERT.
\vspace{0.5cm}
\begin{lstlisting}[frame=single,caption=Plugin Insertion - Old Version ,label=lst:plugInsert1]
<build>
 <plugins>
 </plugins>
</build>
\end{lstlisting}
\vspace{0.5cm}
\begin{lstlisting}[frame=single,caption=Plugin Insertion - New Version ,label=lst:plugInsert2]
<build>
 <plugins>
  (*@\textcolor{red}{<\textbf{plugin}>}@*)
   (*@\textcolor{red}{<\textbf{groupId}>org.apache.maven.plugins</\textbf{groupId}>}@*)
   (*@\textcolor{red}{<\textbf{artifactId}>maven-jar-plugin</\textbf{artifactId}>}@*)
   (*@\textcolor{red}{<\textbf{version}>2.6</\textbf{version}>}@*)
  (*@\textcolor{red}{</\textbf{plugin}>}@*)
 </plugins>
</build>
\end{lstlisting}
\vspace{-0.1cm}
\subsection{Approach}
\label{ss:approach}

This section presents our \differname{} approach to extract \numberofchangetypes{} types of changes from \maven{} build files. 
Our approach is mainly motivated and inspired by the work of Gall \etal \cite{Gall2009ChangeAnalysis} and Fluri \etal \cite{Fluri2007ChangeDistilling}, who showed that detailed information on source code changes can aid in understanding the evolution of software projects, and the work of Macho \etal \cite{Macho2016Predicting}, who showed that this information can be used for  computing models to predict when build configurations should be updated. 

Concerning changes in build configuration files, in particular \maven{} build files, the finest level of analysis that has been performed was on line level. D\'{e}sarmeaux \etal \cite{desarmeaux2016mavenLifecyclePhases} mapped lines of a \maven{} \texttt{pom.xml} to the respective build lifecycle phase. To the best of our knowledge, we are the first to present an approach to extract changes in \maven{} build files on the level of \maven{} configuration elements, that we refer to as fine-grained build changes.

\differname{} first reads two versions of a \maven{} build file (\ie \texttt{pom.xml}) and represents each version as a tree. Then, it uses the  GumTree \cite{Falleri2014GumTree} implementation of Dotzler \etal  \cite{Dotzler2016moveoptimizeddiff} to extract the differences between the two trees in terms of edit operations to transform one tree into the other. The list of edit operations is then mapped to the \numberofchangetypes{} change types that are defined in our taxonomy. In the following, we present each step in detail:

\textbf{Preprocess Build Files.}
The first step of \differname{} preprocesses the two versions of a \maven{} build file. 
\maven{} build files are descriptive, meaning that the order of the elements in the file can be changed without changing its semantics. We observed that GumTree can match elements of the same level more accurately if they are sorted. Hence, \differname{} first sorts the elements on the same level according to their content.\footnote{Strings are sorted alphabetically and numbers in ascending order} For example, the tag \texttt{<module>MySubmodule</module>} appears before \texttt{<module>TheModule</module>}.
Furthermore, \differname{} removes comments and attributes. Attributes, such as \texttt{combine.children} and \texttt{combine.self} for plugin configuration inheritance, affect the build configuration at execution time. We only analyze the build configuration from a static point of view and hence, we remove attributes.

\textbf{Extract Edit Operations.}
Next, \differname{} parses the two preprocessed versions of a \maven{} build file into two trees and passes them to the GumTree differencing algorithm. 
GumTree provides a TreeGenerator for XML files. Unfortunately, this implementation does not handle values of tags in XML documents. Therefore, we implemented our own TreeGenerator that transforms XML files into GumTree trees. We use the prominent Java XML library \textit{jdom} to read the XML file, and methods provided by GumTree to create the tree. 

GumTree then uses a \texttt{Matcher} instance to find mappings between two trees. \differname{} extends the GumTree's default matcher by adding a mechanism to ensure that only tags with the same name will be matched, and by modifying the similarity calculation of two nodes. Tags that have a child tag named \texttt{id} are matched if the Levenshtein similarity of the \texttt{id} value exceeds a threshold $t$. The matcher chooses the node with the highest similarity exceeding the threshold. Tags that have the \maven{} triplet (\texttt{groupId}, \texttt{artifactId}, \texttt{version}) as child nodes are matched by applying the Levenshtein distance for \texttt{groupId} and \texttt{artifactId}. Two nodes are matched if the Levenshtein similarity exceeds a threshold $t$. The matcher chooses the node with the highest similarity exceeding the threshold. Experiments with different $t$ values suggest that $t=0.65$ yields the best performance.

Given the matcher, GumTree outputs a list of tree edit operations comprising added, deleted, updated, and moved elements in the tree that transform the source tree (previous version of the \maven{} build file) into the target tree (subsequent version of the \maven{} build file). 

\textbf{Sort Edit Operations.}
\differname{} considers a particular order to process the changes in \maven{} files. 
We process the operations of the edit script in a top down order according to their level in the build file (parent nodes first). \differname{} applies this order to prevent the extraction of additional changes that result from the insertion and deletion of composite \maven{} tags that also insert or delete their children at the same time. For instance, when a new dependency is inserted, \differname{} only records a DEPENDENCY\_INSERT, skipping the insertion of the child tags of that dependency (\eg \texttt{groupId}, \texttt{artifactId}, and \texttt{version}).

\textbf{Map Build Changes.}
In this step, \differname{} maps the tree edit operations that are generated by GumTree to the \numberofchangetypes{} change types of our taxonomy. 
We consider insertions, deletions, and updates. We do not consider moves, since \maven{} build files are descriptive, meaning that the order of the elements in the file can be changed without changing its semantics. 

To map edit operations to change types, \differname{} iterates over the sorted edit operations mapping each edit operation to at most one change type. Changes to child elements are handled by first checking whether the change is part of an insertion, deletion, or update of its parent. In that case, the change to the child element is not mapped, since it is already part of the parent change. For instance, the insertion of a dependency is mapped only to the change DEPENDENCY\_INSERT while the insertions of its child elements \texttt{groupId}, \texttt{artifactId}, and \texttt{version} are skipped.

As a result, \differname{} outputs a list of build changes that have been performed between two versions of a \maven{} file. 


\section{Evaluating \differname{}}
\label{sec:dataPreparation}
In this section, we describe the evaluation of \differname{}. First, we describe how we selected the projects and how we extracted the data that we used for the evaluation of \differname{} and the experiments. Second, we present the evaluation of our prototype implementation of the \differname{} approach in terms of precision and recall. Finally, we discuss examples of correctly and incorrectly extracted changes.

\subsection{Data Preparation}
The projects that we selected for our experiments stem from two origins. First, we selected the list of projects that Macho \etal used in their prior study \cite{Macho2016Predicting} because they cover a wide range of different vendors, project sizes, and purposes. Second, we extended this list with popular projects from GitHub to improve the variety of the selected projects. We retrieved a list of Java projects ordered by their star rating\footnote{as of January 2017} and removed projects that do not use \maven{} as build system and projects with less than 3500 commits in the repository or rated with less than 1000 stars. We calculated a ranking metric by adding the number of commits and the star rating to evenly balance the user rating and the number of commits of a project. Then, the list of projects was sorted according to the ranking value in a descending order. From this list, we selected the top 20 projects and added them to our list of projects. \tabref{tab:projectData} shows the full list of selected projects and several descriptive statistics of commits and build changes.

\begin{table*}[ht!]
	\renewcommand{\arraystretch}{1.0}
	\caption{List of Java projects used for evaluating \differname{} and for studying the evolution of build files plus descriptive statistics of BCC (Commits with Build Change), BCCR (ratio of BCC), NBCC (Commits without Build Change), NBCR (ratio of NBCC), BC (Build Changes), and \#R (Number of extracted Releases)}
	\label{tab:projectData}
	\centering
	\resizebox{\textwidth}{!}{%
		\begin{tabular}{|l|l||r|r||r|r|r|r|r||r|r|r|r|r||r|}
		\hline
		Vendor & Name & Rank & Stars & \#Commits & \#BCC & BCCR & \#NBCC & NBCR &  BC & \#R \\ 
		\hline
 \hline
neo4j & neo4j & 49,663 & 3,344 & 46,319 & 9,684 & 0.21 & 36,635 & 0.79 & 78,551 & 170 \\ 
hazelcast & hazelcast & 24,995 & 1,850 & 23,145 & 1,306 & 0.06 & 21,839 & 0.94 & 7,914 & 126 \\ 
SonarSource & sonarqube & 23,989 & 1,431 & 22,558 & 1,746 & 0.08 & 20,812 & 0.92 & 18,332 & 109 \\ 
Alluxio & alluxio & 23,545 & 2,787 & 20,758 & 1,694 & 0.08 & 19,064 & 0.92 & 12,973 & 25 \\ 
languagetool-org & languagetool & 21,949 & 1,021 & 20,928 & 285 & 0.01 & 20,643 & 0.99 & 5,027 & 17 \\ 
netty & netty & 21,443 & 8,637 & 12,806 & 1,395 & 0.11 & 11,411 & 0.89 & 7,791 & 160 \\ 
orientechnologies & orientdb & 20,458 & 2,769 & 17,689 & 1,326 & 0.07 & 16,363 & 0.93 & 8,866 & 94 \\ 
spring-projects & spring-boot & 20,356 & 9,312 & 11,044 & 2,922 & 0.26 & 8,122 & 0.74 & 29,699 & 68 \\ 
h2oai & h2o-2 & 19,071 & 2,116 & 16,955 & 74 & 0.00 & 16,881 & 1.00 & 194 & 324 \\ 
google & guava & 18,643 & 13,676 & 4,967 & 221 & 0.04 & 4,746 & 0.96 & 850 & 64 \\ 
deeplearning4j & deeplearning4j & 16,891 & 5,044 & 11,847 & 1,097 & 0.09 & 10,750 & 0.91 & 7,454 & 42 \\ 
stanfordnlp & CoreNLP & 16,688 & 2,685 & 14,003 & 847 & 0.06 & 13,156 & 0.94 & 3,365 & 0 \\ 
eclipse & jetty & 16,357 & 1,148 & 15,209 & 2,965 & 0.19 & 12,244 & 0.81 & 82,153 & 272 \\ 
Graylog2 & graylog2-server & 15,822 & 2,600 & 13,222 & 1,465 & 0.11 & 11,757 & 0.89 & 5,464 & 141 \\ 
prestodb & presto & 15,052 & 5,455 & 9,597 & 1,069 & 0.11 & 8,528 & 0.89 & 10,832 & 178 \\ 
apache & storm & 14,055 & 3,791 & 10,264 & 1,244 & 0.12 & 9,020 & 0.88 & 9,945 & 25 \\ 
apache & flink & 13,466 & 1,869 & 11,597 & 1,343 & 0.12 & 10,254 & 0.88 & 11,022 & 25 \\ 
druid-io & druid & 12,427 & 4,215 & 8,212 & 1,947 & 0.24 & 6,265 & 0.76 & 20,329 & 391 \\ 
naver & pinpoint & 11,976 & 2,864 & 9,112 & 1,012 & 0.11 & 8,100 & 0.89 & 5,950 & 11 \\ 
google & closure-compiler & 11,899 & 2,948 & 8,951 & 60 & 0.01 & 8,891 & 0.99 & 168 & 75 \\ 
apache & activemq & - & - & 11,135 & 1,754 & 0.16 & 9,381 & 0.84 & 12,988 & 51 \\ 
apache & camel &  - & -  & 35,649 & 8,003 & 0.22 & 27,646 & 0.78 & 110,150 & 107 \\ 
apache & hadoop &  - & -  & 48,582 & 2,395 & 0.05 & 46,187 & 0.95 & 25,057 & 240 \\ 
apache & hbase &  - & -  & 29,097 & 2,161 & 0.07 & 26,936 & 0.93 & 10,059 & 538 \\ 
apache & karaf &  - & -  & 15,953 & 5,853 & 0.37 & 10,100 & 0.63 & 54,655 & 60 \\ 
apache & wicket &  - & -  & 31,456 & 1,527 & 0.05 & 29,929 & 0.95 & 13,322 & 243 \\ 
hibernate & hibernate-search &  - & -  & 5,976 & 1,177 & 0.20 & 4,799 & 0.80 & 6,251 & 105 \\ 
jenkinsci & jenkins &  - & -  & 26,286 & 4,551 & 0.17 & 21,735 & 0.83 & 28,138 & 483 \\ 
spring-projects & spring-roo &  - & -  & 6,440 & 675 & 0.10 & 5,765 & 0.90 & 10,173 & 35 \\ 
wildfly & wildfly &  - & -  & 23,370 & 5,186 & 0.22 & 18,184 & 0.78 & 43,384 & 74 \\ \hline
& Sum     &  - & -  & 543,127 & 66,984 & - & 476,143 & - & 641,056  & 4,253 \\ 
& Average &  - & -  &  18,104 &  2,233 & 0.12 &  15,871 & 0.88  & 21,369   & 141.77 \\ 
 \hline
		\end{tabular}%
	}
\end{table*}

For each project, we extracted the build changes as follows: First, we cloned the repository and iterated over each commit, including commits on branches. Second, we checked for modifications in \maven{} build files (\texttt{pom.xml}) indicated by Git. For each of the modified build files, we determined its preceding version and provided both versions to \differname{} to extract \maven{} build changes. The extracted changes were stored in a database called the ChangeDB. Due to its size, we provide the database for other researchers on request. If a commit was a merging commit (\eg pull requests that were merged into the master branch), we did not extract the changes of this commit because we already extracted these changes in the respective commits of the branch. 

\subsection{Evaluation}
\newcommand{\numberofinstances}{400}
\newcommand{\numberofjunittests}{250}
\newcommand{\numberofmissingchangetypes}{37}
We evaluated \differname{} in two ways: first, we used extensive JUnit testing to check whether our approach can detect each single build change type. Second, we performed a manual evaluation with \numberofinstances{} build-changing commits that were randomly selected from the build files of the \numberofprojects{} open source Java projects.

\subsubsection{JUnit Tests}
For each build change type, we first constructed a \texttt{pom.xml} pair that contains exactly one instance of a particular change type and the corresponding JUnit test case. 
In addition, we selected subsequent versions of actual \texttt{pom.xml} files from open source Java projects that contain multiple changes and created JUnit tests for those as well. These pairs have been selected by an expert with more than 7 years of \maven{} experience to cover changes that are often performed in practice. In total, we developed \numberofjunittests{} JUnit tests showing that \differname{} is capable of correctly extracting and classifying each single build change in isolation, as well as co-occurring build changes. 

\subsubsection{Manual Investigation}
To show that \differname{} is also working on real world projects and that it can extract build changes as they are understood by software developers, we invited two PhD students to evaluate \numberofinstances{} build-changing commits. Both students are studying software engineering and have more than 4 years of experience with \maven{}. They received the same set of \numberofinstances{}  \texttt{pom.xml} pairs containing 745 build changes. Prior to the experiment, we briefly explained our taxonomy of build changes to them. In the experiment, we asked each student to label the changes in the \texttt{pom.xml} pairs according to our change taxonomy and compared the output among the participants and with the output of \differname{}.

\textbf{Data Selection.}
\tabref{tab:projectData} shows the list of open source Java projects from which we randomly selected 
\numberofinstances{} commits that contained changes to a \texttt{pom.xml} build file. 
We calculated the sample size based on a population size of 66,984 commits that contain build changes (\tabref{tab:projectData}, sum of column \#BCC), with a margin of error of 5\% and a confidence level of 95\%. The minimum sample size is 382 commits\footnote{\url{https://www.checkmarket.com/sample-size-calculator/}} and we finally decided to randomly select \numberofinstances{} commits to exceed the minimum sample size. 

The total amount of build changes in the data set is 641,056. By randomly selecting the commits for the evaluation, we missed to evaluate  \numberofmissingchangetypes{} of the \numberofchangetypes{} change types (or 44\%). However, the analysis of these missing change types in our subject systems showed that these changes only represent 0.9\% of the total build changes (5,760 out of 641,056). Hence, we can safely assume that our sample set sufficiently covers the majority of build changes in our data set. 

\textbf{Evaluation.} For each of the \numberofinstances{} commits, we provided the study participants with the original and modified version of the build file. We asked each participant to go through all of the selected commits and assign each change to the corresponding change type from our taxonomy. We then compared the changes that were assigned by the participants with the list of build changes extracted by \differname{}.
With the results from each participant, we calculated \textit{precision} and \textit{recall} to measure the performance of our approach. Following the approach of Fluri \etal \cite{Fluri2007ChangeDistilling} used for evaluating ChangeDistiller, we calculated precision and recall as:

\[\textrm{precision} = \frac{\textrm{\#relevant changes found}}{\textrm{\#changes found}}\]
\[\textrm{recall} = \frac{\textrm{\#relevant changes found}}{\textrm{\#changes expected}}\]

\textit{Precision} measures how many of the changes that were extracted by our approach were also detected by a study participant. \textit{Recall} measures how many of the changes that a study participant has found have also been found by our approach. Similar to the evaluation of Dintzner \etal \cite{Dintzner2013Extracting}, we are able to evaluate the correctness and the completeness of our approach with these performance measures. 
The results of the evaluation show a high precision and recall of 0.9513 and 0.9796, respectively for Participant 1. For Participant 2, the results show a precision of 0.9601 and a recall of 0.9844. Averaging the values of both participants, we obtain a mean precision of 0.9557 and a mean recall of 0.9820. 
The detailed results of the evaluation are provided in the supplementary material.\textsuperscript{\ref{fn:repPackage}}

We also evaluated the errors per build change type according to our taxonomy. We used the classification of the two manual evaluations to calculate precision and recall per change type. 
We found that 36 out of the 58 change types could be detected among the randomly selected evaluation commits with precision and recall of 1. This group contains changes, such as DEPENDENCY\_VERSION\_UPDATE and GENERAL\_PROPERTY\_INSERT. 9 change types showed a precision and recall between 0.80 and 1. The other 13 change types that were contained in the evaluation scored lower than 0.80, \eg PLUGIN\_UPDATE. We found that especially PLUGIN\_UPDATE is a change type that is difficult to detect properly by our approach, because it is tightly coupled with PLUGIN\_CONFIGURATION\_UPDATE.

Besides the quantitative evaluation, we performed a qualitative evaluation to find out in which scenarios \differname{} shows a good performance and in which it does not. We present an example where \differname{} did not achieve a proper change extraction compared to a human evaluation. We refer the interested reader to the supplementary material\textsuperscript{\ref{fn:repPackage}} to find more types of wrong classification.
%
The example is taken from the flink project\footnote{\url{http://goo.gl/rWPFDy}} where a dependency definition was changed. \lstref{lst:example1old} shows the dependency in the old version and \lstref{lst:example1new} shows the updated version of the dependency. \differname{} extracted two changes, DEPENDENCY\_DELETE and DEPENDENCY\_INSERT. In fact, this is an update of the same dependency where the \texttt{groupId} and the \texttt{artifactId} changed simultaneously. Hence, the correct classification would be a DEPENDENCY\_UPDATE. Our approach could not detect this change correctly because we use a distance measure to match the nodes of two build files. In this case, the measure indicated that the two dependency definitions are not close enough to be considered the same dependency and consequently, \differname{} extracted the two changes wrongly.

\begin{lstlisting}[frame=single,caption=Example of an incorrect Classification - Old Version ,label=lst:example1old]
<dependency>
 <groupId>com.(*@\textcolor{red}{typesafe.akka}@*)</groupId>
 <artifactId>akka-testkit_${scala.binary.version}</artifactId>
 <scope>test</scope>
</dependency>
\end{lstlisting}
\begin{lstlisting}[frame=single,caption=Example of an incorrect Classification - New Version ,label=lst:example1new]
<dependency>
 <groupId>com.(*@\textcolor{red}{data-artisans}@*)</groupId>
 <artifactId>(*@\textcolor{red}{fl}@*)akka-testkit_${scala.binary.version}</artifactId>
 <scope>test</scope>
</dependency>
\end{lstlisting}

In conclusion, we observe that:\\

\noindent\fbox{\begin{minipage}{0.97\columnwidth}
		\differname{} is capable of extracting changes from \maven{} build files with an average precision of 0.96 and an average recall of 0.98. DEPENDENCY\_\-VERSION\_\-UPDATE and GENERAL\_\-PROPERTY\_\-INSERT are among the change types that achieve the best performance, whereas PLUGIN\_UPDATE is among the change types with the highest rate of error.
	\end{minipage}}\\


\section{Build Change Frequency (RQ1)}
\label{sec:changeFrequency}
Our first experiment investigates the frequency of build changes. We aim to gain knowledge of which change types are frequently performed in projects and how often they are performed. This information can help to understand the evolution of build files similar to the study of Gall \etal \cite{Gall2009ChangeAnalysis}. With this experiment, we aim to answer RQ1: \textit{\RQOne}

\textbf{Approach.} First, we checked the projects for their number of build changes. The projects \texttt{h2o-2} (last commit: Nov 4, 2014), \texttt{guava} (last commit: Jan 1, 2017, only 11 build changing commits in 2016), and \texttt{closure-compiler} (last commit: Nov 15, 2016, only 4 build changing commits in 2016) contain less than 1000 build changes each. Hence, we excluded them from the experiment because we assume that they do not use \maven{} actively and only keep the \maven{} configuration in the source code management system for legacy reasons. 

Second, starting with the change data stored in the ChangeDB, we iterated over the remaining 27 projects and counted the occurring changes. We counted the number of each change type per project and aggregated the numbers also per change category. As depicted in \tabref{tab:projectData}, the selected projects differ in their size, and hence, contain a different amount of total build changes (column BC). Given this variance in the number of build changes, we normalized the change counts to allow a fair comparison between the studied projects. We divided each aggregated change count by the number of total build changes in the project. For example, project \texttt{spring-roo} contains a total amount of 9,534 build changes and 222 instances of the change type MODULE\_INSERT. We calculated the relative occurrence of this change type with $222 / 9534 = 2.33\%$ and used this relative value instead of the absolute value for our experiment. 
We then analyzed this data in two ways. First, we analyzed the relative occurrence of each build change type, and second, we analyzed the number of build changes per change type category. 


\textbf{Results.} 
The most frequently occurring change type is PARENT\_\-VERSION\_\-UPDATE with a relative frequency of 0.41 on average, meaning that on average 41\% of the build changes are of this type. The second most occurring change type is PROJECT\_\-VERSION\_\-UPDATE having an average relative frequency of 0.08. We observe a large drop (0.33) in the relative frequency of those two types of changes underlining that PARENT\_\-VERSION\_\-UPDATE is the most frequently occurring change type by far. These two change types are followed by DEPENDENCY\_\-INSERT (0.06) and GENERAL\_\-PROPERTY\_\-UPDATE (0.03). 

\begin{figure}[!ht]
	\centering
	\includegraphics[width=\columnwidth]{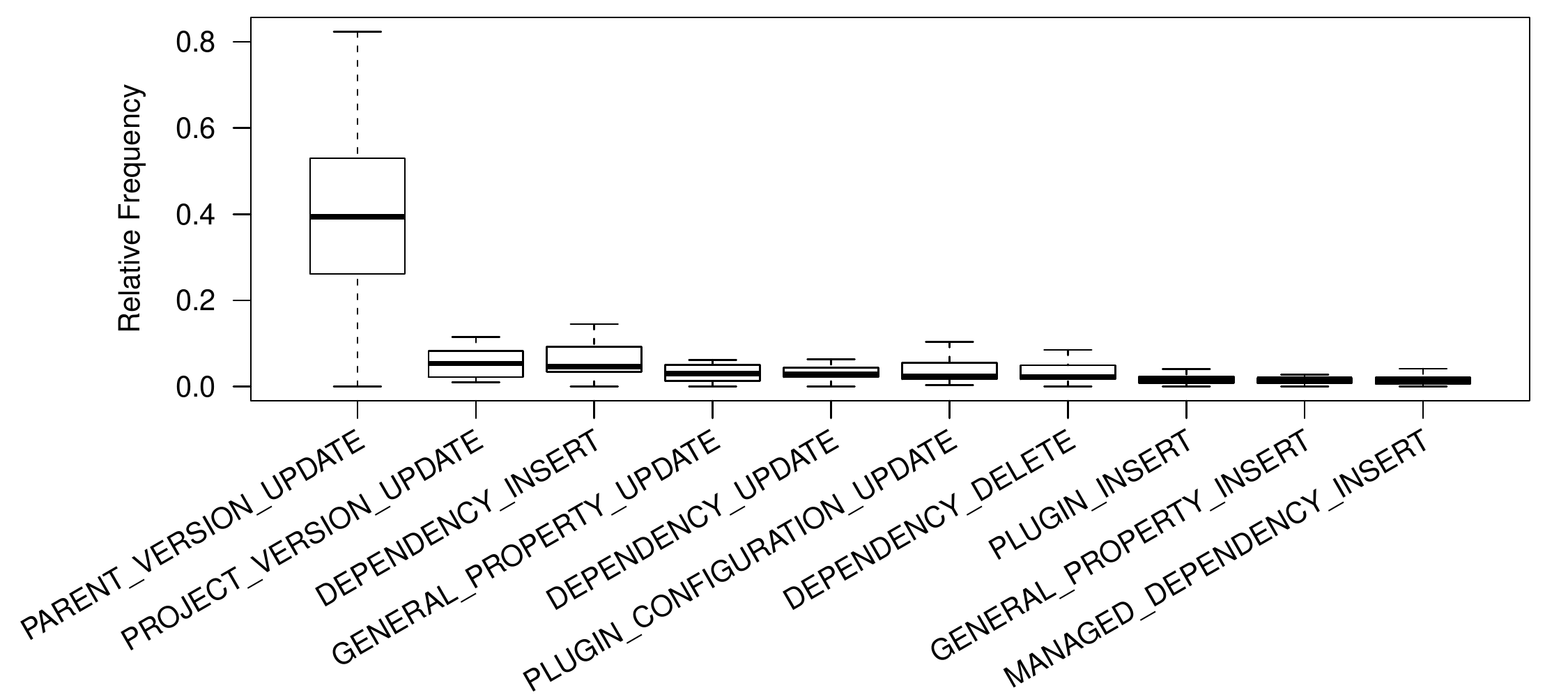}
	\caption{Boxplots of the relative build change frequency for the top 10 most frequent change types (sorted by median).}
	\label{fig:changeFrequencies}
\end{figure}

We see that the top 10 most frequent change types consist of four change types that modify dependencies, and two change types concerning version changes, plugin changes and changes to properties, respectively. This indicates that the dependency management system, which is a core part of the build system, is changed frequently. We also observe that the configurations of plugins are frequently changed which indicates the importance of this part of the build configuration. Furthermore, we observe that the top 10 change types account for 73\% of all changes. \figref{fig:changeFrequencies} shows boxplots of the top 10 most frequent change types.

\begin{figure}[!tbh]
	\centering
	\includegraphics[width=\columnwidth]{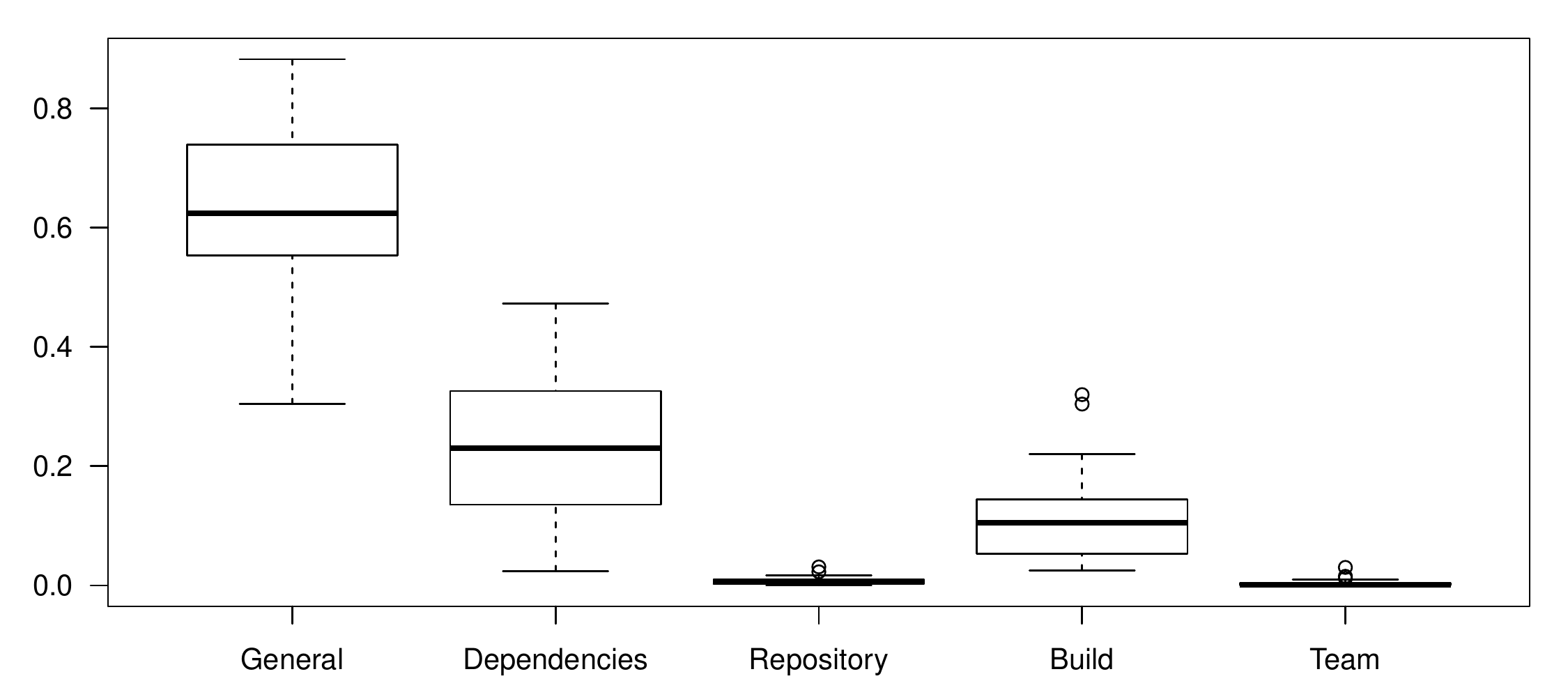}
	\caption{Boxplots of the relative frequency of build changes per change type category.}
	\label{fig:catFrequencies}
\end{figure}

The next step of the analysis deals with the frequencies of build changes per change category that we have defined in \secref{ss:taxonomy}. \figref{fig:catFrequencies} shows the relative frequencies of the build changes per build category. We observe that the \texttt{General Changes} category accounts for 0.64 (64\%) of all changes on average. We argue that this ratio is as expected because changes to the properties, parent changes, and changes to the project metadata, such as project version, are aggregated in this category. 

Furthermore, we can see that \texttt{Dependency Changes} are the second most frequent change category (0.24). This is in line with the observations of the single change types. As mentioned above, the dependency management system is a core part of the build system and is frequently updated. The third most frequently occurring category contains the changes to the \texttt{Build Changes} category (0.11). Lastly, changes to the \texttt{Repository Changes} and to the \texttt{Team Changes} are rare (0.008 and 0.004, respectively). 

%

With these results, we can answer research question RQ1:\\

\noindent\fbox{\begin{minipage}{0.97\columnwidth}
		Among the top 10 most frequently occurring change types, we find version changes and dependency changes frequently. The most frequent change type is PARENT\_VERSION\_UPDATE followed by PROJECT\_\-VERSION\_\-UPDATE, and DEPENDENCY\_\-INSERT. The most frequently occurring change category is \texttt{General Changes} directly followed by \texttt{Dependency Changes}, and \texttt{Build Changes}.
	\end{minipage}}\\



\section{When are the changes recorded (RQ2)}
\label{sec:whenAndWhy}
In this section, we investigate when the build changes occur. We suppose that build changes are not equally distributed over the project, but have phases in which they occur significantly more frequently than in other phases. Hence, we used the build change data that we extracted using \differname{} to check whether our hypothesis holds and answer RQ2: \textit{\RQTwo}

\textbf{Approach.}
We started with the aggregated change data that we created in \secref{sec:changeFrequency}. This data contains for each commit of a project the number of changes per change type that have been performed in the commit. For this research question, we added the date on which the commit was performed and summed up all build changes to a single value per day, \ie one row of our dataset contains the ID of the commit, the number of build changes that were performed in that commit, and the date of the commit. Based on this information, we investigated the data in two ways, as a single day value and with a sliding window approach. 

The first investigation treats each day as a single data point, and hence, adds the number of build changes of commits that were made on the same day. For example, if exactly two commits were made on $23^{th}$ June 2016 with 10 and 15 build changes, respectively, we created a single data point with 25 build changes. The second investigation uses this data and applies a sliding window approach, similar to the approach of Maarek \etal \cite{maarek1991slidingwindow}. We summed the number of build changes of $k$ days to increase the context of the build changes. 

As our hypothesis for this research question states, we suppose that build changes are not equally distributed over the project, but occur more frequently in some time periods of the project. 
We further suppose that one special period in the project that shows a significantly higher amount of build changes, is the time around releases. Thus, we extracted the release data of each of the studied projects provided by the GitHub API. In particular, we extracted the commit ID of the release, the day of the release, and its name. Column \#R of \tabref{tab:projectData} shows the number of releases per project that we could extract. We can see that for the project \texttt{CoreNLP}, we could not retrieve release data. Hence, we also excluded this project and performed the experiment with the remaining 26 projects.

To substantiate our claim, we will show that days that are close to a release contain statistically significantly more build changes than days that are not close to a release. We consider a single day as well as a sliding window approach. To that extent, we consider a day to be close to a release if it is in between $k$ days before the release. For the analysis on a daily basis, we consider $k=1$ and for the sliding window approach, we consider $k\in\{5,7,9\}$ days. We choose different values for $k$ to investigate the influence of the size of the window on the results. We did not try with larger window sizes because we argue that changes that happen more than 9 days before a release should not be considered close to a release. This argumentation is in line with the release data that we used because the average number of days between two consecutive releases is 14 days. Hence, selecting larger $k$ values would possibly cover more than one release. We performed the study with all of the $k$ values and the results were similar. Thus, we only present results for $k=7$ in the paper. The results for $k\in\{5,9\}$ can be found in the supplementary material.\textsuperscript{\ref{fn:repPackage}}

Next, we checked if the distributions are significantly different with a Mann-Whitney-Wilcoxon test ($\alpha<0.01$) and calculated the effect size $d$ using Cliff's Delta \cite{cliff1993dominance}. We used Mann-Whitney-Wilcoxon and Cliff's Delta since the number of build changes is non-normal distributed. The effect size is considered negligible for $d < 0.147$, small for $0.147 \leq d < 0.33$, medium for $0.33 \leq d < 0.47$, and large for $d \geq 0.47$ \cite{Grissom2005EffectSize}. 

%

\begin{figure*}[!t]
	\centering
	\includegraphics[width=\textwidth]{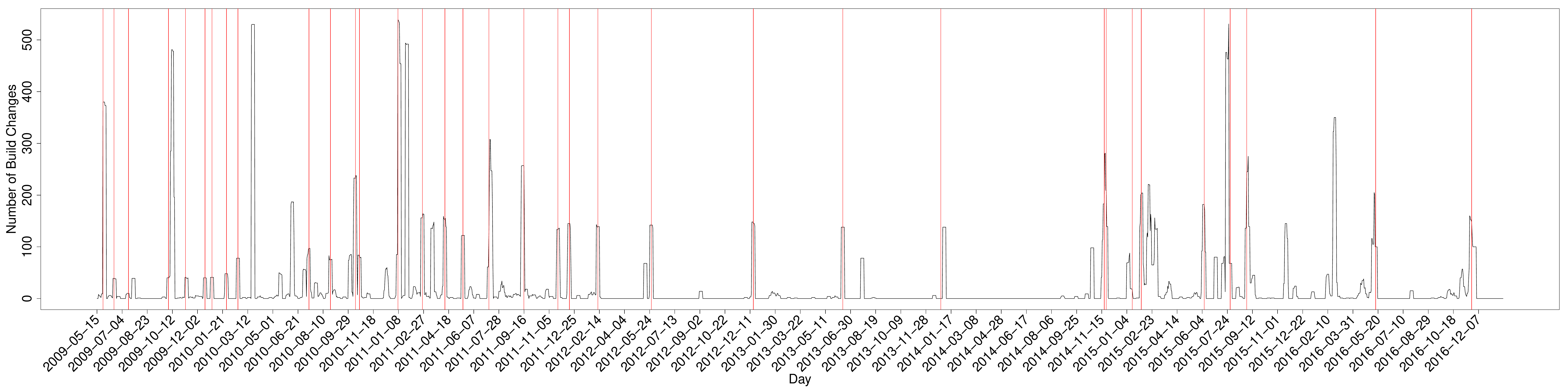}
	\caption{Number of build changes over time in \texttt{spring-roo} using a sliding window ($k=7$). Releases are depicted as vertical red lines.}
	\label{fig:numberOfBuildChangesSliding}
\end{figure*}

\textbf{Results.}
\figref{fig:numberOfBuildChangesSliding} shows the distribution of build changes across the project \texttt{spring-roo}. We refer the reader to the supplementary material\textsuperscript{\ref{fn:repPackage}} for the figures of all of the studied projects. The black line depicts the number of build changes according to the sliding window approach ($k=7$). Each vertical red line indicates a release. We can see that most of the peaks of the black line (number of build changes) appear near a vertical red line (release). This suggests that our hypothesis is correct. 

%
\begin{figure}[!t]
	\centering
	\includegraphics[width=\columnwidth]{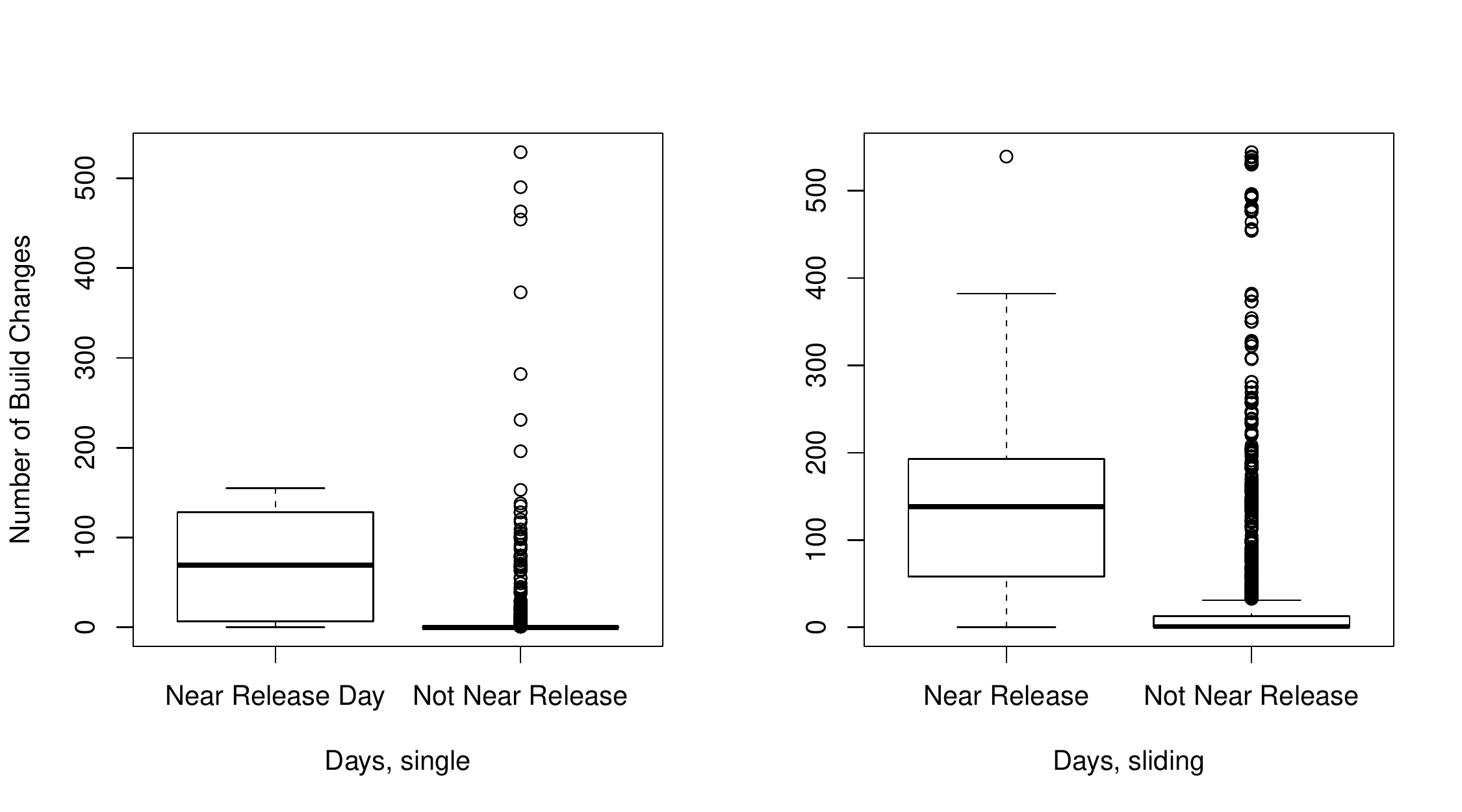}
	\caption{Boxplots showing the distributions of the counts of near release and non-near release build changes of \texttt{spring-roo} as computed with the single day approach and sliding window approach using $k=7$.}
	\label{fig:buildChangeDistribution}
\end{figure}

Furthermore, looking at the distribution of the number of build changes in days near releases and comparing it with the distribution of build changes in days that are far from releases, we can see that the distributions appear to be different. \figref{fig:buildChangeDistribution} shows the boxplots for both approaches and both distributions. The left two boxplots represent the single day approach whereas the right two boxplots represent the sliding window approach using $k=7$. In each of the two plots, the respective boxplot on the left depicts the distribution of build changes on days near a release whereas the respective boxplot on the right depicts the distribution of build changes on days that are not near a release.


\tabref{tab:testAndEffectSize} presents the p-value (p) of the Mann-Whitney-Wilcoxon test and Cliff's Delta $d$ for each project and approach in detail. The p-values show that the frequency of build changes near and not near a release differ significantly (all $p < 0.01$). Furthermore, the effect size can be considered large in all projects except \texttt{hadoop} (small) and \texttt{wicket} (medium). The \texttt{jenkins} project achieves a large effect size with the single day approach but only shows a small effect size with the sliding window approach. The \texttt{presto} project also shows a large effect size with the single day approach but only a medium effect size with the sliding window approach. We find that these lower effect sizes are caused by the release information. \texttt{hadoop} and \texttt{wicket} show a dense release plan in the beginning of the data and this can influence the sliding window approach. Similarly, the changes of the \texttt{jenkins} and \texttt{presto} projects are performed on the release day and hence, the inclusion of additional days, as done by the sliding window approach, lowers the effect size.

\begin{table}[]
	\renewcommand{\arraystretch}{1.0}
	\caption{Results of the Mann-Whitney-Wilcoxon test (p: p-value) and Cliffs Delta $d$ of the distributions of the number of build changes near and non-near releases using the single day approach and sliding window approach with $k=7$.}
	\label{tab:testAndEffectSize}
	\centering
	\begin{tabular}{|l|rr|rr|}
		\hline
		\multirow{2}{*}{Project} & \multicolumn{2}{c|}{Single Day} & \multicolumn{2}{c|}{Sliding ($k=7$)} \\\cline{2-5}
		& p & d & p & d \\ 
		\hline
		activemq & 2.73E-35 & 0.75 & 3.31E-15 & 0.64 \\ 
		alluxio & 3.17E-15 & 0.79 & 7.93E-06 & 0.52 \\ 
		camel & 3.03E-57 & 0.89 & 1.20E-29 & 0.64 \\ 
		deeplearning4j & 1.33E-34 & 0.96 & 2.00E-17 & 0.77 \\ 
		druid & 8.57E-225 & 0.97 & 2.42E-92 & 0.68 \\ 
		flink & 8.78E-21 & 0.89 & 3.59E-10 & 0.73 \\ 
		graylog2-server & 2.47E-117 & 0.84 & 3.21E-21 & 0.47 \\ 
		hadoop & 1.43E-26 & 0.30 & 1.21E-16 & 0.30 \\ 
		hazelcast & 1.53E-58 & 0.61 & 1.86E-25 & 0.54 \\ 
		hbase & 1.49E-194 & 0.63 & 2.38E-102 & 0.57 \\ 
		hibernate-search & 1.14E-177 & 0.98 & 3.53E-42 & 0.76 \\ 
		jenkins & 5.28E-235 & 0.86 & 5.10E-08 & 0.15 \\ 
		jetty & 2.17E-200 & 0.96 & 8.73E-72 & 0.66 \\ 
		karaf & 1.79E-48 & 0.98 & 1.34E-26 & 0.80 \\ 
		languagetool & 2.26E-99 & 0.94 & 3.27E-24 & 0.94 \\ 
		neo4j & 2.07E-68 & 0.70 & 1.23E-34 & 0.55 \\ 
		netty & 1.55E-203 & 0.94 & 2.05E-50 & 0.69 \\ 
		orientdb & 8.71E-93 & 0.94 & 2.68E-28 & 0.66 \\ 
		pinpoint & 1.94E-14 & 0.95 & 6.59E-07 & 0.86 \\ 
		presto & 1.27E-145 & 0.98 & 2.07E-23 & 0.46 \\ 
		sonarqube & 8.37E-85 & 0.93 & 1.46E-25 & 0.59 \\ 
		spring-boot & 2.81E-50 & 0.97 & 3.45E-26 & 0.76 \\ 
		spring-roo & 8.27E-42 & 0.78 & 8.57E-17 & 0.77 \\ 
		storm & 9.55E-15 & 0.70 & 4.20E-06 & 0.53 \\ 
		wicket & 1.19E-59 & 0.37 & 3.94E-29 & 0.41 \\ 
		wildfly & 2.61E-45 & 0.93 & 9.97E-26 & 0.71 \\  
		\hline
	\end{tabular}%
\end{table}

With these results, we can answer research question RQ2 as follows:\\

\noindent\fbox{\begin{minipage}{0.97\columnwidth}
		Build changes are not equally distributed over the projects' timeline. There are particular phases which show significantly higher build change frequencies than others. Especially around releases, a high build change frequency is observed.
	\end{minipage}}\\

\section{Discussion}
\label{sec:discussion}
In this section, we first discuss a number of implications of our results on recent and ongoing research of build systems and their configuration. Next, we discuss implications for developers who use \maven{} as build system. Finally, we discuss the threats to the validity of our results.

\subsection{Implications of the Results}
\textbf{On Research.}
First, compared to the state-of-the-art our fine-grained build changes enable a more detailed analysis of the co-evolution of source code and build files. Second, studies on effort estimation, such as that from Sarro \etal \cite{sarro2016multi}, can be refined by taking into account our \numberofchangetypes{} types of build changes. Third, refactoring approaches, such as MAKAO \cite{Adams2007DesignRecovery} and Formiga \cite{Hardt2015EmpiricalFormiga} can be applied to \maven{} build files and enriched with our detailed change information to improve the refactoring process of \maven{} build files. Fourth, studies of build complexity \cite{McIntosh2012Evolution} can also benefit from our detailed analysis of build changes by including dynamical information, such as our detailed build change information, to the calculation of the metrics. Finally, our build changes can be used to improve the models to predict bug-prone build files \cite{Giger2011ComparingBugPrediction} or suggest potentially missing changes to build configurations \cite{Macho2016Predicting} that might lead to a build breakage.

\textbf{On Development.} 
We observed that build changes occur more frequently near releases. This observation can help developers to avoid build breakage by increasing the awareness that each change to the build configuration can possible break the build. Furthermore, project managers can use this finding to consider the peak of build changes near releases in their planning of releases and work load. We also give insight into the type of build changes that are frequently made. This can be used by developers, for instance, to identify and refactor plugins that often change their configuration.

\subsection{Threats to Validity}
\label{sec:threats}
Regarding the validity of our results, we identified the following threats to construct, internal, and external validity.

\textbf{Construct Validity.}
One threat is that our taxonomy may not cover all possible change types or change categories that could be theoretically made to a build file. We mitigated this threat in two ways. First, we compared the taxonomy with the XML schema of \maven{} build files to cover all important changes. Second, we asked two experienced \maven{} users to verify the taxonomy and create the categories, including a discussion if necessary. 
Furthermore, we retrieved the release data of the 30 open source Java projects from GitHub as the only resource. To that extent, we could miss possible releases if they are not covered by the GitHub data. However, we mitigated this threat by manually checking if the data is compliant with the data provided by the source code management system. 

\textbf{Internal Validity.}
A threat to internal validity concerns whether \differname{} can extract the changes to a build configuration file accurately. We mitigated this threat by covering all changes of the taxonomy with JUnit tests and a manual evaluation comparing against the opinions of two experienced \maven{} users. 
Concerning the evaluation of \differname{}, a threat is that the randomly selected commits do not include all change types. We mitigated this threat by calculating the proportion of actually missed changes due to the selection. We observed that we only miss 0.9\% of the changes and hence, we can safely assume that the majority of the changes will be covered by \differname{}.


\textbf{External Validity.}
The main threat to external validity stems from the selection of projects that we used in our study. We mitigated this threat by selecting 30 open source Java projects of different vendors, sizes, and purposes. However, additional experiments with projects using other build systems and from industrial settings are needed to further generalize our results. Another threat to external validity is that our taxonomy is tailored for \maven{} build configurations. While we designed the taxonomy to be usable for other build tools as well, the taxonomy may not generalize to all other build systems.

%
%
%

\section{Conclusions}
\label{sec:conclusions}
Build systems are an essential part in the engineering process of modern software systems. 
In this paper we introduced \differname{}, an approach for extracting fine-grained build changes from \maven{} build files. In a manual evaluation, we showed  
that \differname{} is capable of extracting build changes with an average precision and recall of 0.96 and 0.98, respectively. 
With the build changes extracted from \numberofprojects{} open source Java projects we performed two empirical studies to investigate the frequency and time of build changes. The results of the two studies showed:
%
\begin{itemize}
	\item (RQ1) The most frequent change type is PARENT\_\-VER\-SION\_\-UP\-DATE followed by PROJECT\_\-VERSION\_\-UPDATE, and DEPENDENCY\_\-INSERT. The most frequent change category is \texttt{General Changes} directly followed by \texttt{Dep\-end\-ency Changes}, and \texttt{Build Chan\-ges}. The top 10 change types account for 73\% of all changes.
	
	\item (RQ2) Build changes are not equally distributed over the projects' timeline. We observed that especially near releases build changes occur more frequently.
		
%
\end{itemize}

Our results benefit research on build configurations and developers using \maven{} as their build system. 

%
%
%

\textbf{Future Work.}
We plan to extend \differname{} to support other build systems, such as Gradle,\footnote{\url{https://gradle.org}} and compare them with \maven{}.
Furthermore, we will investigate build changes to find frequent change patterns among commits and work items that affect the build result. Finally, we plan to perform a more detailed analysis of the co-evolution between build changes and source code changes.
\bibliographystyle{IEEEtran}
\bibliography{IEEEabrv,references}
%
%
%

\end{document}